\documentclass[a4paper,fleqn,usenatbib,useAMS]{mnras}

\usepackage[dvips]{graphicx}
\usepackage{color}
\usepackage{times}
\usepackage{natbib}
\usepackage{setspace}
\usepackage{amsmath}
\usepackage{amsfonts}
\usepackage{amssymb}
\usepackage{multirow}
\usepackage{multicol}
\usepackage{aas}
\usepackage{booktabs}
\usepackage{adjustbox}
\usepackage{float}
\usepackage{graphics}

\usepackage[applemac]{inputenc}
\def\lav{\langle}
\def\rav{\rangle}

\def\II{_{\rm II}}
\def\beq{\begin{equation}}
\def\eeq{\end{equation}}
\def\se{^{\rm e}}
\def\der{{\rm d}}
\def\modot{M$_\odot$}

\title[Ionised bubble at z=6.5]{An ionised super-bubble powered by a proto-cluster at z=6.5}
\author[J.M.~Rodr\'i{}guez Espinosa et al.]{J.M. Rodr\'i{}guez Espinosa$^{1,2},$\thanks{E-mail: jre@iac.es} J.M. Mas-Hesse$^3$, E. Salvador-Sol\'e$^4$,  
\newauthor R. Calvi$^{1,2}$, A. Manrique$^4$, K. Chanchaiworawit$^{5}$,  
\newauthor R. Guzman$^{5}$, J. Gallego$^6$, A. Herrero$^{1,2}$, and A. Mar\'in Franch$^7$  \\
$^1$Instituto de Astrof\'isica de Canarias, E-38205 La Laguna, Spain \\ 
$^2$Depto. de Astrof\'isica, Universidad de La Laguna, E-38206 La Laguna, Spain\\
$^3$Centro de Astrobiolog\'ia (CSIC-INTA), Depto. de Astrof\'isica, Madrid, Spain \\
$^4$Institut de Ciencies del Cosmos, Universitat de Barcelona, UB-IEEC. Mart\'i Franqu\'es 1, E-08028 Barcelona, Spain\\
$^5$Dept. of Astronomy, University of Florida, Gainesville, USA\\
$^6$Depto. de Astrof\'isica y CC de la Atm\'osfera, Universidad Complutense de Madrid, Spain\\ 
$^7$Centro de Estudios de F\'isica del Cosmos de Arag\'on (CEFCA) -  Plaza San Juan, 1, E-44001, Teruel, Spain  
}

\pubyear{2020}

\begin{document}

\label{firstpage}
\pagerange{\pageref{firstpage}--\pageref{lastpage}}
\maketitle

\begin{abstract}
  We show herein that a proto-cluster of Ly$\alpha$ emitting galaxies, spectroscopically confirmed at redshift 6.5, produces a remarkable number of ionising continuum photons. We start from the Ly$\alpha$ fluxes measured in the spectra of the sources detected spectroscopically. From these fluxes we derive the ionising emissivity of continuum photons of the proto-cluster, which we compare with the ionising emissivity required to reionise the proto-cluster volume. We find that the sources in the proto-cluster are capable of ionising a large bubble, indeed larger than the volume occupied by the proto-cluster. For various calculations we have used the model AMIGA, in particular to derive the emissivity of the Lyman continuum photons required to maintain the observed volume ionised. Besides, we have assumed the ionising photons escape fraction given by AMIGA at this redshift.

\end{abstract}


\section{Introduction}
Reionisation followed the dark ages when enough Population III stars and galaxies were in place that their ionising output was sufficient for the task. Population III stars and star-forming galaxies started the reionisation process by forming primordial bubbles of ionised gas. These bubbles grew, illuminated by galaxies with strong ionising Lyman continuum, and merged, till the entire Universe became ionised. High-z Ly$\alpha$ emitters are perhaps the most important witnesses and key players of the re-ionisation process. Indeed, at redshifts larger than 6, low luminosity Ly$\alpha$ sources would not be visible, unless they are located within sizeable bubbles of ionised gas.

The last decades have witnessed a surge in studies of the cosmic history of the Universe. In particular the reionisation of the Universe has received much attention, though it is a process not yet completely understood.  Population III stars and star-forming galaxies started the re-ionisation process by $z \sim 30$. Reionisation proceeded by forming primordial bubbles of ionised gas. Bubbles that grew and merged till through percolation the Universe became ionised. Beyond $z\geq 3$, the best strategy to find distant galaxies and proto-clusters  \citep{Naidu2018,Kakiichi2018,Harikane2019,Higuchi2019} is to look for the so-called Ly$\alpha$ emitters (LAEs), or the prominent rest-frame UV continuum galaxies, also known as Lyman Break galaxies (LBGs). Furthermore, there is increasing evidence that low luminosity star-forming galaxies were the main culprits for reionising the Universe \citep{Ouchi09,Bouwens10,Robertson15,Finkelstein2012}.

The most distant, overdensity discovered so far is at $z\approx 7$ \citep{Castellano2018}. These authors did find an over-density in the Bremer Deep Field, which they claim to be ionised. Other medium and high-z proto-clusters have been reported in recent years \citep{Oteo2018, Abdullah2018, Jiang2018}, such as those discovered at $ z \sim 5.7$ and $6.6 $ by \citet{Higuchi2019,Harikane2019}. Finally, while this work was being refereed, a paper was published showing evidences of a bubble ionised by 3 Lyman alpha emitting galaxies at z = 7.7 \citep{Tilvi2020}. 

This paper deals with the ionisation state of a proto-cluster at redshift 6.5, discovered by us \citep{Kritt17, Chanchaiworawit2019, Calvi2019}. In particular, in \citet{Calvi2019} we show the spectroscopic confirmation of a sample of the proto-cluster sources. Moreover, we claim that the mechanical energy thrown out by supernovae and stellar winds in the proto-cluster is huge ~\citep{Calvi2019}. Thus we expect that this mechanical energy will pierce holes in the circumgalactic medium (CGM) throughout which Lyman continuum photons would be able to escape to the intergalactic medium (IGM), hence ionising it. 

A key point in this paper is the derivation of the total emissivity of Lyman continuum photons in the observed volume, as well as the required value to keep it ionised. For these calculations we have used AMIGA (Analytic Model of Intergalactic-medium and Galaxies \citep{Manrique2015,Salvador-Sole2017}, a very complete model of galaxy formation which is able to recover all the observed cosmic properties at high-$z$ and meets the observational constraints on re-ionisation drawn from the Cosmic Microwave Background anisotropies \citep{Salvador-Sole2017}. In particular we have assumed the ionising photon escape fraction and luminosity function considered by this model at this redshift.

The paper uses the Ly$\alpha$ escape fractions from \citet{Chanchaiworawit2019} Section~\ref{new}. Then in Section~\ref{defs} we establish a set of definitions, which are meant to set up the grounds for the rest of the paper. Section~\ref{correction} corrects for the number of sources in the mask. We continue with Section~\ref{Phot_protocl} where we compute the ionising emissivity, produced by the proto-cluster, which is available to ionise the IGM. Finally, in Section~\ref{Phot_required} we evaluate the ionising emissivity necessary to ionise the overdensity. We end up with the conclusions.

\begin{table*}
	\begin{tabular}{@{}ccccccc}
    \hline
    Source & $f_{{esc,Ly}\alpha}$ & $L_{\alpha,{intr.}}$ & $Q_{ion}$ & $Q^*_{ion}$ & $\dot {N}_{ion}$ \\
     &  & $ 10^{42}\, erg\, s^{-1}$ & $10^{54}\, s^{-1}$ & $10^{54}\, s^{-1}$ & $10^{52}\, s^{-1}$ \\
    \hline
    C1-01 & 0.19$\pm$0.10 & 78.95$\pm$41.55 & 6.62 $\pm 3.49$ & $6.99 \pm 3.68$ & $37.07 \pm 1.96$\\
    C1-02 & 0.17 $\pm 0.02$ & 17.65 $\pm 5.92$ & 1.48 $\pm 0.50$ & $ 1.56 \pm 0.52 $ & $8.29 \pm 0.44$ \\
    C1-05 & 0.20$\pm$0.09 & 6.00 $\pm$3.08 & 0.50  $\pm 0.26$ & $ 0.53 \pm 0.27 $ & $2.82 \pm 0.15$\\  
    C1-11 & 0.18$\pm0.09$ & 15.56$\pm$ 8.49 & 1.31 $\pm 0.71$ & $ 1.38 \pm 0.75 $ & $ 7.30 \pm 0.39$\\
    C1-13 & 0.20$\pm$ 0.10 & 25.50$\pm$ 13.73 & 2.14  $\pm 1.15$ & $ 2.26 \pm 1.22 $ & $11.97 \pm 0.64$ \\
    C1-15 & 0.19 $\pm$0.07 & 12.11 $\pm$5.03 & 1.02  $\pm 0.42$ & $ 1.07 \pm 0.45 $ & $5.68 \pm 0.30$\\
    C2-20 & 0.18$\pm$0.10 & 7.78$\pm$4.68 & 0.65 $\pm 0.39$ & $ 0.69 \pm 0.41 $ & $4.65 \pm 0.19$ \\
    C2-29 & 0.18$\pm$0.08 & 18.33$\pm$8.97 & 1.54 $\pm 0.75$ & $1.62 \pm 0.80 $ & $8.61 \pm 0.46$\\
    C2-35 & 0.26$\pm$0.14 & 6.15 $\pm$4.85 &  0.52 $\pm 0.41 $ & $ 0.55 \pm 0.43$ & $2.89 \pm 0.15$\\
    C2-40 & 0.19$\pm$0.07 & 21.58$\pm$8.77 & 1.81$\pm$0.74 & $1.91 \pm$ 0.78 & $10.13 \pm 0.54$\\
    C2-43 & 0.19$\pm$0.08 & 5.26$\pm$2.43 & 0.44 $\pm 0.20 $ & $0.47 \pm 0.22 $ & $2.47 \pm 0.13$\\
    C2-46 & 0.20$\pm$0.07 & 12.00$\pm$5.22 & 1.01 $\pm 0.44 $ & $1.06 \pm 0.46 $ & $5.63 \pm 0.30$ \\
    Sum  & & & $38.85 \pm 4.02$ & $41.02 \pm 4.24$ & $ 217.37 \pm 2.32$ \\
    \hline
    \end{tabular}
    \caption{columns: 1) Name; 2) Ly$\alpha$ escape fraction 3) Intrinsic Ly$_{\alpha}$ Luminosity; 4) Effective number of ionising continuum photons per second; 5) intrinsic or ``stellar" number of photons per second (assuming f$_{{esc,LyC}}=0.053$); (6) Number of ionising continuum photons escaping from the galaxy and eventually participating in reionising the intergalactic medium. Note that the last row is the sum of sources C1-05 till C2-46 multiplied by the completeness factor (2.8125) plus the two Ouchi sources C1-01 and C1-02. Errors have been propagated quadratically.} 

\label{Table1}
\end{table*}

\section{The data}
\label{new}
We have used the Ly$\alpha$ Luminosities from  \citet{Calvi2019}. Besides, we have taken the Ly$\alpha$ escape fractions from \citet{Chanchaiworawit2019}. We would like to mention that the values we have used are consistent with previous works in the SXDS field, which estimated the Lyman alpha escape fraction at $z \approx 6.5$ at $0.30 \pm 0.18$ \citep{Ouchi10}.

\section{Definitions}
\label{defs}
To clarify a few concepts that will be used throughout the paper we define them here:
\begin{itemize}
\item Intrinsic Ly$\alpha$ luminosity ($L_{Ly\alpha,intr}$): This is the Ly$\alpha$ luminosity emitted by the local ionised nebula itself, before any absorption or scattering effects. It is derived from the observed Ly$\alpha$ luminosity \citep{Calvi2019} by dividing it by the Ly$\alpha$ escape fraction (f$_{{esc,Ly}\alpha}$). The values of the f$_{{esc,Ly}\alpha}$ were taken from \citet{Chanchaiworawit2019}.  The f$_{{esc,Ly}\alpha}$ and the Ly$\alpha$ luminosities are listed in Table~\ref{Table1}. 

We have taken the intrinsic Ly$\alpha$ luminosities ($L_{Ly\alpha,intr}$) from the 10 sources detected spectroscopically in \citet{Calvi2019}. We have also added two sources already known from the literature \citep{Ouchi10} in the same field. However, while C1-01 was in the spectroscopy mask, thus its Ly$\alpha$ luminosity was derived similarly \citep{Calvi2019}, the C1-02 was not. For this reason we have taken its Ly$\alpha$ luminosity  from \citet{Ouchi10}. 

\item Effective number of ionising continuum photons per second (Q$_{ion}$). This is the rate of ionising continuum photons that corresponds to the intrinsic Ly$\alpha$ luminosity in a typical HII region.   

\item Intrinsic number of stellar ionising continuum photons per second (Q$^*_{ion}$). This is the rate of ionising continuum photons directly emitted by the massive stars, not considering any absorption, scattering or escape of photons from the nebula without effectively ionising the gas. They are derived from the number of effective ionising continuum photons (Q$_{ion}$) dividing it by (1 - f$_{{esc,LyC}}$), where f$_{{esc,LyC}}$ is the Lyman continuum escape fraction. Note that the f$_{{esc,LyC}}$ is different from f$_{{esc,Ly}\alpha}$.

\item Finally, $\dot {N}_{ion}$ is the number of ionising continuum photons that actually participate in reionising the intergalactic medium (IGM). $\dot {N}_{ion}$ is derived multiplying Q$^*_{ion}$ by f$_{esc,LyC}$, which has been assumed to be 0.053 (see the discussion below).

\end{itemize}

\section{Correction for the number of sources in the mask}
\label{correction}
The sources that we were able to insert in the observing mask were only 16 out of the 45 Ly$\alpha$ emitter candidates detected photometrically \citep{Kritt17}, in the Subaru/XMM-Newton Deep Survey field. Out of these 16 sources, we had reliable spectra for only 10. Note that these sources were selected for being the brighter. But also to include as many sources as possible without the spectra falling on top of each other. In fact, the average Ly${\alpha}$ luminosity of the 45 candidate sources is $6.87 \times\,10^{42}$ erg $s^{-1}$, while the average from the 10 sources for which we had spectra is $6.77\times\,10^{42}$ erg s$^{-1}$. Thus the galaxies we detected had a similar average Ly${\alpha}$ luminosity as the whole sample. 
 
Given the number of spectroscopic detections, if we were to have enough slits to accommodate all of the candidates (45) we assume that we would have detected a similar fractional number of sources. Thus we need to include a correction factor to account for those sources not present in the mask. Assuming the same fraction of detections for the 45 candidates, we get $10/16 \times 45 = 28.125$. Since we detected 10, the correction factor is 2.81. The last row in Table~\ref{Table1} is computed as follows: we added together sources C1-05 till C2-46. That sum is multiplied by 2.81, and then, we added the two sources (C1-01 \& C1-02) that were previously known \citep{Ouchi10}.

\section{Ionising photon fluxes produced by the proto-cluster}
\label{Phot_protocl}

Although observing the ionising Lyman continuum flux in high redshift sources is far from the reach of current facilities \citep{Madau1995,Shapley2006,Siana2007}, the effective ionising continuum photon rate (Q$_{ion}$) can be derived from the intrinsic Ly$\alpha$ luminosity, using the relation 
\begin{equation}
   L_{\rm Ly{\alpha}_{intr}} = 1.19 \times 10^{-11}Q_{ion}\,\, {erg}\, {s}^{-1} 
   \label{eq1}
\end{equation}
 
The above equation is based on Case B recombination with T= 10$^4$K and density of $\approx 500$ cm$^{-3}$ \citep{Oti&Mass_Hesse2010}. From this we can compute the effective number of ionising continuum photons, that is the number of Lyman continuum photons (wavelength below 912~\AA) emitted per unit time that are not absorbed by dust and do not escape from the galaxy. Therefore, these photons participate effectively in ionising the inter-stellar medium. The effective ionising continuum photon fluxes, Q$_{ion}$, are also given in Table~\ref{Table1}. 

We are ready now to derive the intrinsic stellar ionising photon flux(Q$^*_{ion}$). This is the total number of ionising continuum photons per unit time, with wavelength shorter than 912~\AA{}, produced by the massive stars. This value has to be necessarily equal to or higher than the ''effective ionising continuum photon flux'' (Q$_{ion}$), since a fraction of this stellar ionising flux escapes from the galaxy, hence it does not participate in the ionisation of the inter-stellar medium, but instead ionises the IGM. Q$^*_{ion}$ is also shown in Table~\ref{Table1}.

The value of the average Lyman continuum photons escape fraction, f$_{{ esc,LyC}}$, has been derived from the results of the AMIGA simulations as discussed in \citet{Salvador-Sole2017}. AMIGA, is a very complete and detailed, self-consistent model of galaxy formation particularly well suited to monitor the intertwined evolution of both luminous sources and the IGM. The multi-parameter fitting to different observables done by AMIGA, constrains the value of the Lyman continuum escape fraction to f$_{{esc,LyC}} = 0.053 \pm 0.007$ at $z = 6.5$. A very similar value, $f_{{esc,LyC}} = 0.05$, has been derived by \citet{Finkelstein2019} using a different methodology. We use the escape fraction obtained from AMIGA, f$_{{esc,LyC}}$, for computing $\dot{N}_{ion}$, whose values are also listed in Table~\ref{Table1}.

Considering the estimated true comoving volume occupied by the protocluster, $V = 11\,410$\, Mpc$^3$ \citep{Chanchaiworawit2019}, we derive an ionising emissivity, from the confirmed and probable LAE candidates in the overdense region, equal to $\dot{{\cal N}}_{ion}=1.91\times 10^{50}$\, phot\, s$^{-1}$ Mpc$^{-3}$.  However those LAEs lie in halos of masses $\ga 10^{11}$ \modot ~\citep{Chanchaiworawit2019}, while, according to AMIGA, galaxies in lower mass halos contribute up to 80\% of the whole ionising emissivity. Thus, the full ionising emissivity in the overdense region should typically be 100/20 times higher, i.e. $\dot{{\cal N}}_{ion}\sim 9.53\times 10^{50}$\, s$^{-1}$\, Mpc$^{-3}$. We want to note that we found \citep{Kritt17} a similar fraction for the luminosity originated in the observed LAEs with respect to the total one, namely $\sim 76\%$. This figure was obtained comparing the number of LAEs observed down to the limiting luminosity of the photometric sample, (L$_{Ly{\alpha}} = 10^{42.4}$\,erg\, s$^{-1}$), with this later value obtained integrating the luminosity function to lower luminosity galaxies.

\section{Number density of photons required to ionise the proto-cluster}
\label{Phot_required}

A non-zero Lyman continuum escape fraction implies that the escaping photons will be able to ionise regions farther out from the galaxy that contains the stars that produce the ionising photons. For instance, let us assume the case of C1-01. As shown in Table~\ref{Table1}, assuming an overall escape fraction of Lyman continuum photons of 5.3\%, this would imply that out from the $6.99 \times 10^{54}$ Lyman continuum photons emitted per second by the massive stars of C1-01, $6.62 \times 10^{54}$ are used in ionising the local interstellar medium (ISM), while the rest,\, $3.70 \times 10^{53}$ photons per second, would escape this individual galaxy becoming available for ionising the IGM.  

 In Section~\ref{Phot_protocl} we derived an ionising emissivity of the overdense region of $\dot {\cal N}_{ion}= 9.53\times 10^{50}\, {s}^{-1}\, {Mpc}^{-3}$. Is such an ionising emissivity enough for the overdense region to be fully ionised? The answer to this question is not straightforward because the ionising photons emitted by all sources in a given volume at one redshift are used to first, balance the recombinations that take place in the ionised bubbles around them, and second, increase the size of those ionised bubbles. More specifically, the equation for the evolving hydrogen ionised fraction, $\dot{Q}\II$, in our overdense volume is the same as for the whole Universe (\citealt{Salvador-Sole2017}; see also \citealt{Manrique2015} for the derivation).
\beq 
\dot Q\II=\frac{{\dot {\cal N}}_{\rm ion}}{{\lav n}\rav}
-\left[\left\lav  \frac{\alpha(T)}{\mu\se}\right\rav\!\frac{C\,\lav n\rav
  }{a^3}+\frac{\der \ln \lav n\rav}{\der t}\right]Q\II
\label{rec}
\eeq 
where $\dot {\cal N}_{ion}$ and the averages in angular brackets are now restricted to that volume instead of to the whole Universe. That is, $\lav n\rav$ is the overdensity (1.67; \citealt{Chanchaiworawit2019}) times the mean cosmic density of hydrogen atoms; $\alpha(T)$ is the temperature-dependent recombination coefficient to neutral hydrogen, where he typical temperature of the ionised bubble in the overdense volume is equal to that of the whole Universe, calculated as explained in \citet{Salvador-Sole2017}, times $1.67 ^{2/3}$ so as to account for the adiabatic contraction produced in the overdensity; $C\sim 3$ is the clumping factor (\citealt{Finkelstein2012}; see also \citealt{Salvador-Sole2017}); $\mu\se$ is the electronic contribution to the mean molecular weight of the cosmic IGM, assuming the usual hydrogenic composition; and $a=0.13$ is the cosmic scale factor at $z=6.5$.

Equation~\ref{rec} shows how, in order to determine the ionised fraction ${ Q}\II$, it is not enough to know the emissivity $\dot N_{ion}$ of the sources in the volume; we also need the rate at which the ionised fraction ${Q}\II$ grows. The AMIGA code provides the mean ionisation fraction $Q\II$ in the Universe at any $z$. However, such a fraction is not known for any particular volume with an arbitrary density. The problem is thus undetermined.

\begin{figure}
\centering
{\includegraphics[scale=0.45]{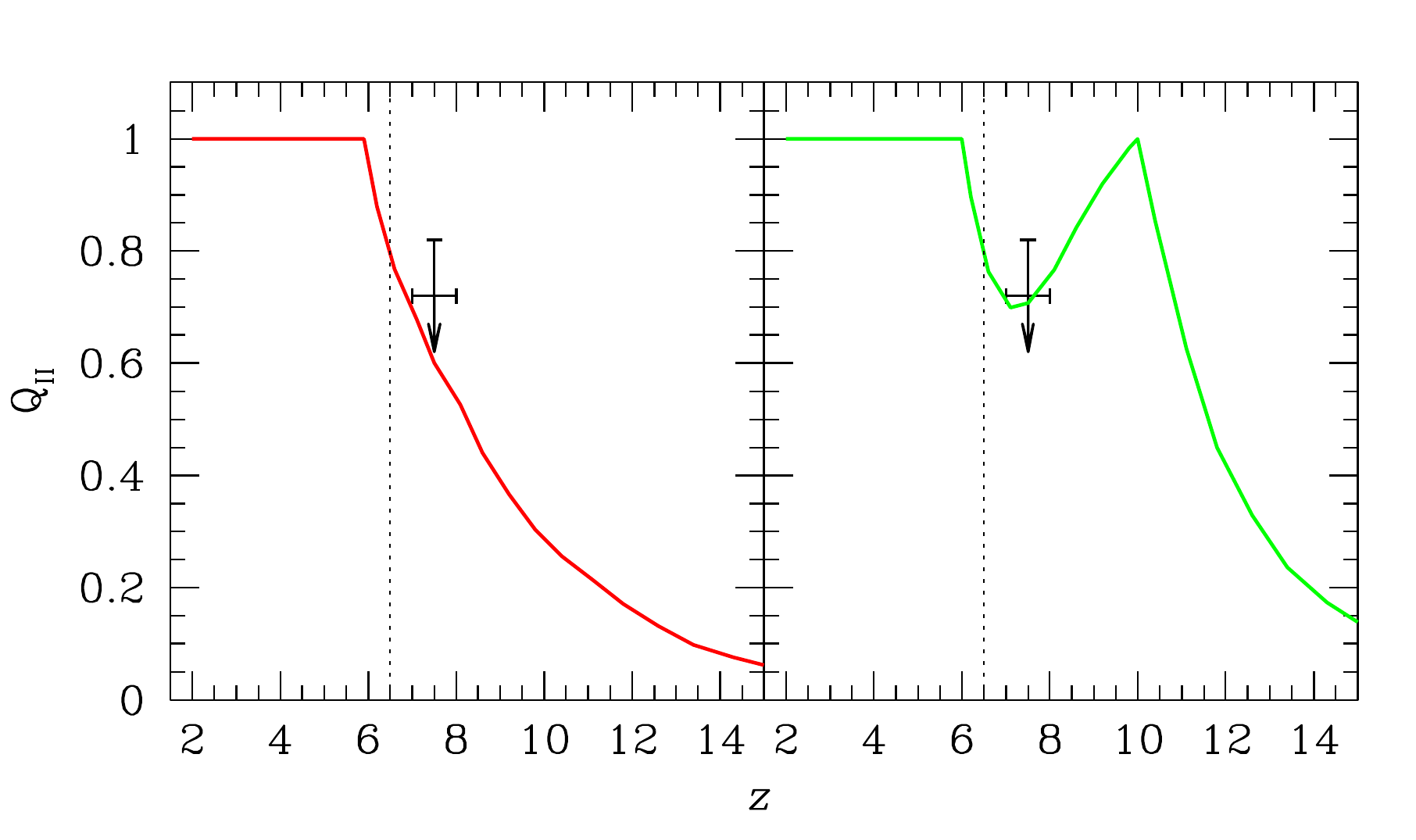}}
 \caption{Evolving average cosmic ionised fraction for the single (left panel) and
   double (right panel) reionisation solutions obtained in \citet{Salvador-Sole2017}. The vertical dotted black line marks the redshift $z=6.5$. The point with error bars gives the upper limit for ${Q}\II$ at $z=7.5$ drawn from the rapid decrease of the LAE abundance at those redshifts \citep{Salvador-Sole2017}. 
   This upper limit puts severe constraints on the ionised fraction at z=6.5, which must be lower than ${Q}\II\sim 0.85$.}
\label{reion}
\end{figure}

Nevertheless, we can circumvent that indeterminacy by asking the minimum emissivity ${\dot{\cal N}}_{ion}^{min}$ needed to just balance the recombinations that would take place in the overdense region, were it fully ionised. The quantity ${\dot{\cal N}}_{ion}^{min}$ can then be drawn from Equation~\ref{rec}, setting $\dot {Q}\II=0$ and ${Q}\II=1$. The result is ${\dot{\cal N}}_{ion}^{min} =4.2\times 10^{50}$\, s$^{-1}$\, Mpc$^{-3}$, i.e. a factor 2 less than the estimated ionising emissivity of the galaxies in the overdense volume, ${\dot{\cal N}}_{ion}= 9.53\times 10^{50}$\, {s}$^{-1}$\, {Mpc}$^{-3}$. Thus, the rate at which ionising photons are produced in that volume is clearly enough to keep it fully ionised and still increase the size of the ionised super-bubble at a rate even higher than the average rate found for the whole Universe at $z=6.5$. This rate can be obtained using Equation~\ref{rec} with a mean cosmic ionised fraction of $Q\II=0.80$ as found by AMIGA at $z=6.5$ for the single reionisation solution (see Fig.~\ref{reion}). Alternatively, the ionised super-bubble could be even larger, even if the rate at which it grows is smaller than the average growth rate of ionising regions in the whole Universe. This latter possibility is, in fact, the most likely because, the small fraction of neutral hydrogen present in the Universe, at $z= 6.5$, should predominantly lie in underdense regions with a lower density of ionising sources. Such underdense regions would need to be ionised more rapidly than average. And  the converse is thus expected for overdense regions.    

The preceding discussion relies on the results of AMIGA for single reionisation. Actually, AMIGA reaches two acceptable solutions, one with single reionisation and the other with double reionisation (see Fig.~\ref{reion}), depending on the initial mass function of Population III stars \citep{Salvador-Sole2017}. However, both solutions are very similar for redshifts around $z=6.5$, so the double reionisation solution leads to essentially the same results. On the other hand, the single reionisation model is most widely accepted so the results given here can be readily compared to those obtained with other ionisation models. The AMIGA model is particularly reliable as it is very complete, self-consistent and monitors the formation and evolution of galaxies and their feedback from trivial initial conditions at the "Dark Ages". The two solutions mentioned are the only ones that satisfy all the observational constraints pertaining to the high-$z$ Universe and the CMB anisotropies. Note that, the independent reionisation model recently published by \citet{Finkelstein2019}, at the redshifts of interest, finds very similar values for all quantities. In particular, they find $f_{esc,LyC}=0.05$ and $Q\II(z=6.5)=0.85$, which are very similar to our own. This fact gives strong support to our calculations.

\section{Conclusions}
\label{conclusions}
We have confirmed spectroscopically a proto-cluster of LAEs in the SXDS/XMM-Newton deep Survey field. For the 10 sources we have spectroscopically confirmed \citep{Calvi2019}, we have  determined their intrinsic Ly$\alpha$ photon luminosities as well as their ionising photon fluxes. We have also derived the corresponding  intrinsic "stellar" ionising continuum photon fluxes. We find that the sources in the proto-cluster produce sufficient Lyman continuum photons to ionise a large bubble. We have done the calculation taking into account the overdensity contrast and the fact that most of the low mass galaxies produce a large fraction of the ionising photon fluxes. For the latter calculation and that of the emissivity of ionising photons required to keep the observed volume ionised, we have relied on the AMIGA model and the Lyman alpha escape fraction derived from it at z$ = 6.5$. Thus, we find that there are sufficient ionising photons to not only ionise the volume occupied by the proto-cluster, but a larger ionised bubble that increases with time. Therefore we claim that we have discovered a large ionised bubble such as those that through percolation completed the re-ionisation of the universe by $z \approx 6$.

\bigskip
Acknowledgements

We want to thank an anonymous referee for his/her helpful comments and suggestions, which helped to significantly improve the manuscript. 
RC, JMRE, ESS \& AM acknowledge support from the Spanish State Research Agency under grant AYA2015-70498-C2-1, and JMRE under grant number AYA2017-84061-P. JMMH is funded by Spanish State Research Agency grants ESP2017-87676-C5-1-R and MDM-2017-0737 (Unidad de Excelencia Mar\'{\i}a de Maeztu CAB). We are indebted to the Severo Ochoa Programme at the IAC. Based on observations made with the Gran Telescopio Canarias (GTC), installed in the Spanish Observatorio del Roque de los Muchachos of the Instituto de Astrof\'\i{}sica de Canarias, in the island of La Palma.

\bibliographystyle{mnras}
\bibliography{ionbub}

\label{lastpage}
\end{document}